\documentclass{interact}

\usepackage{titlesec}
\usepackage{graphicx}
\usepackage[version=3]{mhchem} 
\usepackage{graphicx}
\usepackage{subcaption}
\usepackage{array,multirow}
\usepackage{amsmath}
\usepackage{todonotes}
\usepackage{hyperref}
\usepackage{soul}
\usepackage{algpseudocode,algorithm,algorithmicx}
\usepackage[utf8x]{inputenc}

\usepackage{fancyhdr}

\begin{document}

\title{Sequence Generation using Deep Recurrent Networks and Embeddings: A study case in music.}

\author{
\name{Sebastian Garcia-Valencia\textsuperscript{a}\thanks{Correspondence: Sebastian Garcia Valencia. Email: sgarci18@eafit.edu.co}, Alejandro Betancourt\textsuperscript{b} and Juan G. Lalinde-Pulido\textsuperscript{a}}
\affil{\textsuperscript{a}Computer Science Department, Universidad EAFIT, Medellin, Colombia; \textsuperscript{b}Digital Department, Ecopetrol, Bogota, Colombia}
}

\maketitle

\begin{abstract}
Automatic generation of sequences has been a highly explored field in the last years. In particular, natural language processing and automatic music composition have gained importance due to the recent advances in machine learning and Neural Networks with intrinsic memory mechanisms such as Recurrent Neural Networks. This paper evaluates different types of memory mechanisms (memory cells) and analyses their performance in the field of music composition. The proposed approach considers music theory concepts such as transposition, and uses data transformations (embeddings) to introduce semantic meaning and improve the quality of the generated melodies. A set of quantitative metrics is presented to evaluate the performance of the proposed architecture automatically, measuring the tonality of the musical compositions. 

\end{abstract}

\begin{keywords}
Music Theory; Recurrent Neural Network; Generative Model; LSTM; NAS; UGRNN; Automatic Evaluation; Embedding
\end{keywords}

\section{Introduction} \label{sec:introduction}

Since music origins, music composition has been considered an activity mainly for musicians. By nature, music composition is a creative task and compromises cognitive skills commonly developed exclusively by humans. These cognitive requirements make music composition one of the most challenging tasks to achieve automatically.

During recent years, academic and computer science community has been interested in the use of algorithms to create musical pieces, but just recently, advanced models started showing signs of creativity similar to those of humans \cite{Colombo2017, Johnson2017, DBLP:journals/corr/BahdanauCB14}. These signs of improvement can be explained by i) Recent revolution in computing power and data availability, which facilitates the automatic training of computer algorithms by using examples ii) Recent advances in Algorithmic Structures allowing to include temporal consistency and memory to the automatically composed pieces.

In the process of developing algorithms capable of composing natural and appealing music, multiple strategies have been explored (e.g. Bayesian Algorithms, Neural Networks, among others). In Particular, recent advances in Recurrent Neural Networks (RNN), including memory mechanisms, show that is possible to generate sequences by generalising patterns hidden in large data sets, and have been applied in areas such as automatic music composition and  Natural Language Processing (NLP). In General, RNN is a family of algorithms based on gated architectures composed by self-contained procedures denoted as memory cells.

The training of the RNN is a process in which the memory cells are automatically tuned by using human-made compositions. Once the training of the network ends, it is intuitive to use their learned capabilities as improvisation machines. This feature has different applications in arts, and in particular, in music, it can be used for live play-answer acts, computer-aided performance, and automated accompaniment. In addition to the algorithmic challenges (the data, the algorithmic structure, and the training procedure) automatic music composition must consider musical concepts such as:

\begin{enumerate}
    \item{ \textbf{Music Representation:} To make learning and automatic music composition feasible, it is necessary to use an appropriate data structure (music representation). In this paper, inspired by recent advances in NLP, the notes are encoded as embeddings. The objective is to move the input data to a vector space where the related outputs are closer. In this case, notes which tend to appear in sequence.}
    \item{ \textbf{Long Term Consistency:} A common problem in sequence generation is that, after some time steps, the generated sequence loses sense and becomes random data, this is commonly known as the problem of no long-term consistency. To alleviate this issue,  a common strategy is to use algorithms with memory mechanisms (memory cells). This paper analyses the performance in automatic music composition of different mechanisms, namely, Long Short Term Memory (LSTM), Neural Architecture Search (NAS) and  Update Gate Recurrent Neural Network (UGRNN).}
    \item{ \textbf{Music Transposition:} Music transposition makes two pieces of music to be perceived in essence the same song despite having almost all notes different. This concept, makes the learning of patterns in music sequences quite particular, giving more importance to the changes between notes than to the notes itself. The proposed approach considers three different strategies to exploit the music transposition concept: i) A control case where the sequences are the notes, ii) A modified case storing the pitch intervals, given the first note, it is possible to reconstruct the original sequence, iii) The last case on which each sequence is augmented to 12 through a shifting process with the purpose of having all notes as starting point.}
    \item{ \textbf{Music Evaluation:} Once the model generates some music sequences, it is necessary to quantify the quality of the composition. We propose three metrics considering music theory indications of tonality. The proposed metrics are used to compare different RNN architectures and the impact of the Music Transposition strategies.}
\end{enumerate}

The remaining part of this paper is organised as follows: chapter \ref{sec:stateofart} introduces existing bibliography about automatic music composition with computer science techniques. Chapter \ref{sec:generatingMelodies} introduces our approach for the RNN architecture, the training phase, and the music evaluation. Chapter \ref{sec:experiments} presents the experiments to understand the selected architectures, the objective function, and finally tune the hyperparameters of the proposed networks (e.g., Number of Layers, Memory Cell). Chapter \ref{sec:results} analyses the models and the generated melodies from the quantitative and musical perspective. Finally, chapter \ref{sec:conlusions}, concludes and provides some future research lines of this work.

\section{State of Art} \label{sec:stateofart}

This chapter summarises the main concepts supporting this work moving from music theory to automatic music composition and machine learning. Section \ref{subsec:musicaltheroy} briefly introduces the main musical concepts, for additional music theory concepts see \cite{amsco2000little}. Section \ref{subsec:representation} presents existing strategies including the proposed musical concepts in Machine Learning. Section \ref{subsec:musicgenerationwithalgort} reviews recent architectures and training strategies. Lastly, section \ref{subsec:musiccompositioneval} introduces common ways to evaluate the quality of the generated pieces.

\subsection{Music Theory} \label{subsec:musicaltheroy}

\begin{figure}[h!]
    \centering
    \includegraphics[width=\textwidth]{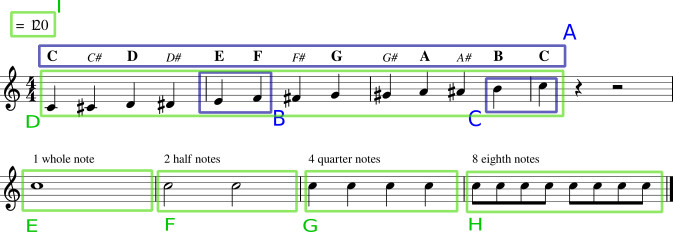}
    \caption{Musical basic concepts}
    \label{fig:musicbasiconcepts}
\end{figure}

A \textbf{note} is the basic element of a musical piece.  A \textbf{melody} is a sequence of notes which sounds at different time steps. There are 12 notes which form the \textbf{chromatic scale}, and are divided into fundamental and altered notes. The fundamental notes (C, D, E, F, G, A, B) form the C major scale and corresponds to the notes of the white keys in a piano (Notes in bold in fig. \ref{fig:musicbasiconcepts}.A). The distance between notes is named \textbf{interval}, being the smallest, the \textbf{semi-tone}, which is the distance between B-C or E-F, see fig. \ref{fig:musicbasiconcepts}.B and \ref{fig:musicbasiconcepts}.C respectively. The semi-tone corresponds to the white keys in a piano next to each other. To represent the altered notes, it is necessary to use \textbf{accidentals} which modify a fundamental note. This modification can be adding (sharp ♯) or subtracting (flat ♭) a semi-tone, these other notes correspond to the black keys in a piano (Notes in italics in fig. \ref{fig:musicbasiconcepts}.A).

As fig. \ref{fig:musicbasiconcepts}.D shows, once the scale reaches B, the next note is again C but in a different position in the sheet stave (second note in fig. \ref{fig:musicbasiconcepts}.C). These groups of 12 notes are known as \textbf{octaves}, the term is used also in psychics, to denote a frequency which is a multiple of other. The frequency of a note, known as \textbf{pitch}, from the next octave will be the double as the corresponding note in the current octave. Figure \ref{fig:musicbasiconcepts}.D begins with the C note of the fourth octave (C4, 261.6Hz) and ends with the C note of the fifth octave (C5, 523.3Hz).

The 12 possible intervals between two notes (including the same note in different octaves) receive the name of \textbf{fundamental intervals}, and once the difference is bigger than 12 semitones, they become \textbf{compound intervals}. In close relation with this concept are the \textbf{tonality} and \textbf{dissonance}. There is a tendency for some intervals to be consonant and perceived pleasant (like 3 semitones or minor third) and others to be dissonant and perceived unpleasant (6 semitones or augmented fourth).

\begin{figure}[h!]
    \centering
    \includegraphics[width=\textwidth]{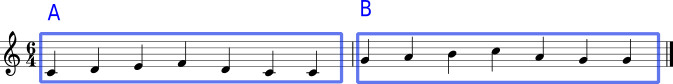}
    \caption{Transposition in music}
    \label{fig:transposition}
\end{figure}

Being explained the previous concepts, it is necessary to introduce \textbf{musical transposition}, which is a characteristic of music that causes that the perceived identity of a piece relates more to the relative changes than to the notes. This feature causes that humans perceive C-D-E-F-D-C-C  (fig.  \ref{fig:transposition}.A)and G-A-B-C-A-G-G (fig. \ref{fig:transposition}.B) as the same melody because they have the same pitch interval series: 2  2  1 -3 -2 0. In section \ref{subsec:dataset} 3 strategies to capture this patterns are presented.

After introducing the pitch, it is necessary to explain \textbf{Time} and \textbf{rythm}. For this matter, there is a set of symbols, the reference value is the whole note (fig. \ref{fig:musicbasiconcepts}.E), equivalent to 2 half notes (fig. \ref{fig:musicbasiconcepts}.F), an all the others are even multiples and subdvisions of it (quarter, eighth and others).

The symbol does not define the duration of the note by itself; it is relative to the \textbf{Tempo} (fig. \ref{fig:musicbasiconcepts}.I). The \textbf{quarter note} (fig. \ref{fig:musicbasiconcepts}.G) is frequently used as reference to define the tempo as the number of quarter notes that sounds in one minute, usually called Beats Per Minute (BPM). For example, in fig. \ref{fig:musicbasiconcepts} all the sections (\ref{fig:musicbasiconcepts}.E, \ref{fig:musicbasiconcepts}.F, \ref{fig:musicbasiconcepts}. G and \ref{fig:musicbasiconcepts}.H) which are known as \textbf{bars} have the same duration of 2 seconds.

This paper only considers as song, melodies of at least 12 notes. All the pieces have a tempo of 120 BPM and contain uniquely quarter notes. The extension to \textbf{harmonies} (which consist of multiple notes at the same time) and time variation is highlighted as future work. 

\subsection{Representation} \label{subsec:representation}

There are different approaches to represent the music for machine learning; however, array-based representations are the most popular and can be divided into direct representations and dimensional transformations. On the one hand, direct representations consider various aspects of the notes or harmonies. The authors in \cite{Hornel1998} define a 12-bit codification over motifs (notes groups) which considers musical knowledge about intervals and harmonic relationships. They also have an independent representation of the harmonies. In \cite{Johnson2017}, the authors use a binary array for the note with the position, pitch class, previous context, beat and previous vicinity. To define vicinity, the authors use an idea taken from convolutional networks, replacing the kernel of the convolution by the RNN, and the pixels by notes of other time steps.

On the other hand, dimensional transformations is an alternative to represent musical data as a more convenient array. Two examples of this approach are the `Dodecaphonic Trace Vector' in \cite{Kaliakatsos-Papakostas2010} and the `token' of \cite{DePrisco2017}. The former uses a normalised vector of musical pieces transposed to the key of C major or A minor; this approach is key tolerant and preserves the general characteristics of the song. The latter is a 3-tuple: k1: the name of the chord, k2: the type of chord (maj7, m7, etc.) and k3: the role of the note concerning the chord. The `Token' approach does not consider time in the representation but as an extra step in the model.

An interesting vector representation called `word embedding' \cite{DBLP:journals/corr/abs-1301-3781} is commonly used in NLP, and is a strategy to transform words in a corpus to points in a vectorial space. Embeddings popularity relies on its ability to locate semantically nearby words in adjacent spacial segments. The context can be as simple as the surrounding words or the actual syntactic context \cite{levy2014}. Embeddings are highlighted as a powerful dimensional reduction and enhance the capability of machine learning algorithms to capture hidden patterns. In practice, the embedding transformation can be learned simultaneously with the RNN.  

Finally, there are also `Non-array' representations such as the `abc language' used in \cite{WalshawChris2017, Oliwa2008}, on which notes and time are represented with chars\footnote{conversion from midi is possible with the abcMIDI package   \url{http://abc.sourceforge.net/abcMIDI/}}.

To the best of our knowledge, the concept of embedding has been only used in music to extract some patterns \cite{sephora2016, herremans2017} but not for automatic composition, which is usually highlighted as future research. In this paper, the n-gram based model \cite{DBLP:journals/corr/MikolovSCCD13} is used to transform the notes and intervals to embeddings, and add semantic meaning to the notes. In \cite{garciavalencia2020embeddings}, we analyse the resulting embeddings. 

\subsection{Music composition with Algorithms} \label{subsec:musicgenerationwithalgort}

Three components are necessary to automatically generate music: the algorithm architecture, the training data and the objective function to optimise.

\textbf{Algortihms and Architectures: } An algorithm is denoted as generative when it can produce information given an input. This input could be a set of numbers which have a given distribution or an array of data containing hidden parameters or weights. Some examples of generative models are Restricted Boltzmann machine \cite{Fischer2012} and Generative adversarial networks \cite{2014arXiv1406.2661G} commonly used for image generation. For automatic sequence generation, the n-gram model \cite{Brown:1992:CNG:176313.176316} is widely used to predict the next element in a sequence probabilistically.

An artificial neural network is a mathematical model inspired by nature. In the biological context, a neuron has three main parts: i) An axon to send signals to other neurons, ii) A dendrite to receive the signals from the axons of others neurons, and iii) The soma, which is the neuron nucleus and is where the signal can be modified \cite{farinella2014neurocomic}. The points where dendrites and axons communicates are called synapses and the way different neurons connect with others in the brain determine what a living being learn, think and do.

In the artificial version, a neuron receives a value and multiplies it by an internal weight before transmitting it to the next neuron or layer of neurons. The organic-dynamic interconnection is simulated by varying the weight of every neuron until it satisfies an error.

The use of generative models for music has been a widespread problem. It has been approached with different strategies such as neural networks \cite{Colombo2017, Hutchings2017, Waite2016, Hornel1998}, Bayes transform \cite{thornton2011}, Bayesian networks with a hidden Markov model \cite{Raphael2002} and probabilistic inference using statistical models \cite{whorley2016, Allan2005}. As a consequence of the recent advances in hardware and data availability, it is becoming popular to use more complex Artificial Neural Networks (ANN) for tasks as computer vision with Convolutional NN (CNN) \cite{Krizhevsky:2012:ICD:2999134.2999257}, which basically make a dimensional reduction of the input layer by layer, and sequence generative models with Recurrent Neural Networks (RNN) \cite{DBLP:journals/corr/Schmidhuber14}. 

Regarding CNNs, different adaptions have been tried, but the absence of memory severely compromises the results. The authors in \cite{2017arXiv170503122G} use a combination of convolutions, gated linear units, and attention mechanisms for language translation. The proposed CNN can not learn sequence positions, as RNN does. The authors alleviate this issue by using positional embeddings. The study concludes that CNNs are capable of discovering compositional structures by building hierarchical representations. However, are less frequent in sequence generation problems. Another interesting approach is proposed in \cite{Johnson2017}, using a two folded architecture, the first part, considers the relationship between notes using convolutions, while the second part considers the temporal consistency. This approach allows the model to learn patterns in both characteristics without losing invariance.

RNNs, in particular, use the output of the network as input in a feedback process. There are different implementations like bidirectional RNN \cite{Berglund2015} which are used to predict lost values in the middle of sequences. A special type of RNN with memory-like characteristics is the gated RNN. These architectures are used in sequence modelling tasks and generative models to address the problem of long-term consistency. In these gated RNN, in addition to the coefficients (neural weights), there are a group of cells or gates that must be trained with the past elements. In this paper three gated architectures are explored: 

\begin{enumerate}
    \item LSTM \cite{Hochreiter1997}: The Long-Short-Term Memory NN is composed of 4 elements: a cell state which determines the current context, a forget gate which decides which information remove, an update gate to insert new information and an output gate to decide the output \cite{Olah2015f}.  
    \item NAS: Neural Architecture Search is a type of cell which designs its internal structure by exploring combinations of recurrent cells dispositions and activation functions. NAS use reinforcement learning with the accuracy as reward \cite{Zoph2016}.
    \item UGRNN: the Update Gate RNN is a variation of a vanilla RNN with a single gate added, which determines if the current state of the cell was updated or taken from a past time step. Authors in \cite{Collins2016} compare it against LSTM and another gated RNN called GRU \cite{2014arXiv1406.1078C}. UGRNN shows to be as good as GRU to learn difficult tasks but easier to train than LSTM and GRU for deep architectures.
\end{enumerate}

The authors in \cite{DBLP:journals/corr/ChungGCB14} compare LSTM and GRU and conclude that GRU performs better in 3 out of 4 datasets. LSTM is used In \cite{Colombo2017} and \cite{Hutchings2017}. The former uses a deep LSTM with 2 different datasets (Irish folk music and Klezmer tunes). The trained algorithm is able to keep the genre during the generation phase. In the later study, an LSTM is the base for real time autonomous music generators. LSTM networks can also extend to other problems like chord estimation \cite{deng2018}.

The magenta project (Google brain) studies two different approaches to address long-term consistency: i) Look-back networks \cite{Waite2016} using custom labels and previous events from 1 and 2 musical bars in the input. ii) Attention networks, improving the long-term memory by weighing the outputs of previous steps. Attention networks were initially used for translation tasks \cite{DBLP:journals/corr/BahdanauCB14}.

\textbf{Training Data: } Music datasets are usually the result of scraping music repositories. These containers can be specific for symbolic music (Gutenberg project \cite{Gutenbergproject2006}, Mutopia project \cite{Mutopiaproject2017}) or general music files (Million song dataset \cite{Bertin-Mahieux2011}). The most popular container in recent years is the MuseScore sheet music archive \cite{MuseScore2002}, because there is a significant community supporting and contributing to it.

Academically there are four open datasets commonly used for polyphonic music (i.e. Nottingham, JSB Chorales, MuseData, Piano midi). These datasets use MIDI format and consist of classical and folk music \cite{2012arXiv1206.6392B}. Other authors have used these datasets to test their models (\cite{Gu2015, DBLP:journals/corr/ChungGCB14}). The authors in \cite{Kaliakatsos-Papakostas2010} use a similar dataset of Piano-midi but with fewer composers. 

\textbf{Objective Function: } Concerning the optimisation of the model in training, the magenta project has an interesting approach based in reinforcement learning \cite{Jaques2016} on which a book of music theory is used to define the rewards and train an LSTM based model called the RL tuner. This approach helps to reduce unpleasant characteristics like notes out of key and improves good aspects like compositions with only one top note.

Another common strategy is to use a defined metric as objective function in a classic optimisation problem. Usually, cross-entropy is a popular choice for classification and generative models \cite{Johnson2017, Hutchings2017, whorley2016, kawthekarevaluating}. According to information theory, the cross-entropy quantifies the differences between two probability distributions \cite{shannon48}. In machine learning, it is used to compare the predicted output with the ground truth, and particularly in generative models, measures the prediction capability \cite{kawthekarevaluating}. Being the ground truth an unknown distribution, the negative average log-likelihood is used as an estimator. Commonly, in NLP the metric chosen is the perplexity, which is the exponentiation of the cross-entropy. For optimisation purposes both are equivalent, but perplexity is more natural to interpret due to its linear scale.

The base architecture for this paper is a multi-cell RNN with two hyperparameters, namely the number of memory cells and their type (i.e. LSTM, NAS, UGRNN). For training purposes we use the mono-midi-transposition-dataset, a monophonic dataset created after applying a number of modifications to the mono-musicxml-dataset proposed in \cite{DBLP:journals/corr/WelU17}, which is the result of scraping musescore for monophonic music, this is an advantage cause the most of the available data tend to be polyphonic. Finally, we use cross-entropy as the objective function. In \cite{garciavalencia2020cross} we validate the use of cross-entropy showing the gradual reduction in the metric loss, reflects a reduction of randomness of the generated melodies. 

\subsection{Music Composition Evaluation} \label{subsec:musiccompositioneval}

According to the recent bibliography, music evaluation can be divided into two groups. The first one uses mathematical formulations \cite{Colombo2017, Jaques2016, Tymoczko2011}, while the second one uses musical experiments to consider human judgments about the quality of the composition. 

The quantitative evaluation of musical compositions defines mathematical formulations to measure the quality of a musical piece according to music theory. The authors in \cite{Colombo2017} introduce a new metric called novelty to quantify the musical creativity. The idea is to use a corpus of the training to calculate how different are the generated songs with respect to training dataset. In general, quantitative metrics such as novelty are useful to judge the quality of musical pieces automatically and teach generative algorithms. However, they ignore the subjective feeling that humans easily detect.

To alleviate this issue, other authors propose human-based experiments. A good example is a test made by \cite{DePrisco2017}, where professional musicians are asked to score musical pieces by answering two questions with a number between 1-7: (1) `How do you rate the quality of music?', and (2) `How do you rate the coherence of the music with the X style?'

Finally, there is also the option of using a combination of both methods. This is the case of the study proposed in \cite{Hutchings2017} which combine perplexity and human experts judgement. Perplexity assesses a part of the model called the harmonic improviser. Then, the generated pieces are scored by professional musicians in 6 dimensions like harmonic consistency, thematic development, and melodic creativity.

This paper uses a combination of cross-entropy for training and quantitative metrics for evaluation of music tonality. See section \ref{subsec:quantmetricdescr} for the mathematical implementation of the metrics

\section{Generating melodies} \label{sec:generatingMelodies}

Following the ideas from section \ref{sec:stateofart}, our goal is to develop a network to generate melodies. Section \ref{subsec:rnnarchitecture} describes the architecture and training of the RNN. Section \ref{subsec:dataset} illustrates the method to scrap the data from musescore, preprocess it, clean it and create the two variations to introduce the transposition principle. Finally, section \ref{subsec:quantmetricdescr} formalises the implementation of the three quantitive metrics and shows the result of applying them to the dataset. This result is used as the baseline for the performance analysis of the results section \ref{subsec:automEval100songs}.
\newline

\subsection{RNN Architecture} \label{subsec:rnnarchitecture}

The base for the proposed architecture of our RNN is char-rnn (\cite{Karpathy2015}), originally introduced to generate text char by char. The main variations with respect original char-rnn are:  i) The embedding encoding and decoding of music, ii) the use of 3 different cell types in the multi-cell RNN.

\begin{figure}[h!]
    \centering
    \includegraphics[width=\textwidth]{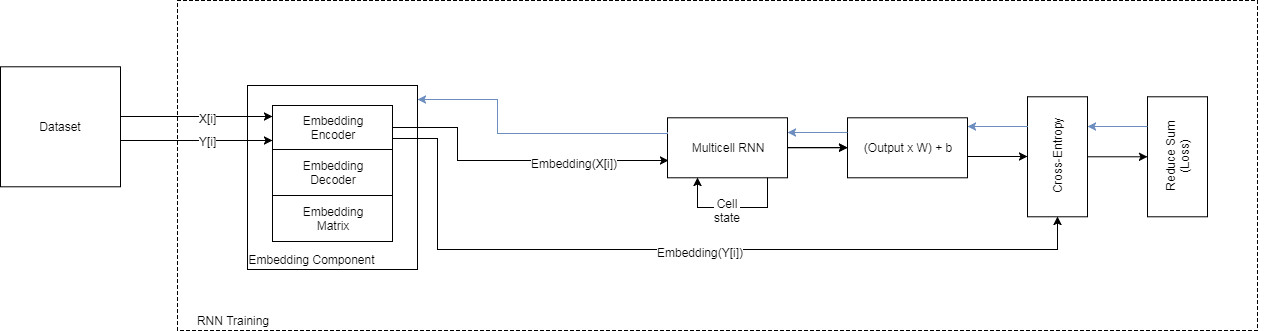}
    \caption{RNN Architecture training mode}
    \label{fig:rnnarchitecturetraining}
\end{figure}

\begin{figure}[h!]
    \centering
    \includegraphics[width=\textwidth]{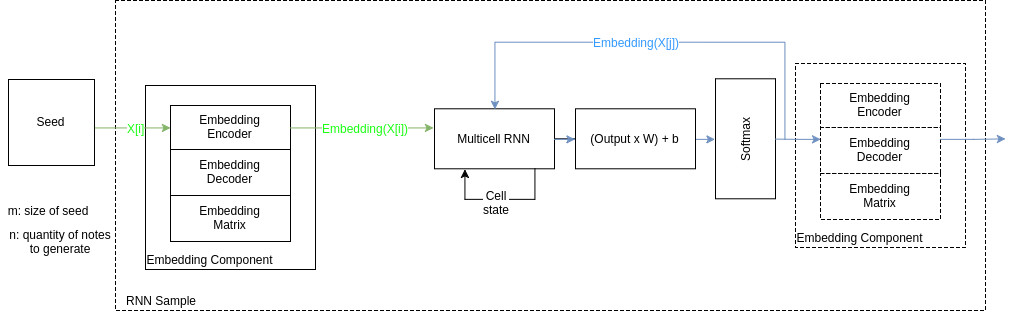}
    \caption{RNN Architecture sampling mode}
    \label{fig:rnnarchitecturesampling}
\end{figure}

The workflow through the network varies depending on the use, namely training mode (fig. \ref{fig:rnnarchitecturetraining}) or sampling mode (fig. \ref{fig:rnnarchitecturesampling}). For training, the starting point is the dataset. As explained in section \ref{subsec:dataset}, the three dataset variants have the same two elements: the X and Y array, the later is the former shifted by one position. First, the embedding component of the network produces a matrix using all the distinct elements in the dataset, the dimension of this matrix will be the size of the vocabulary multiplied by the embedding dimension. Then, every sample goes first to the embedding component to become a vector.

The embedding goes to the multicell RNN; this is the main component of the network which can contain an array of cells of any type (LSTM, NAS or UGRNN). Once the embedding flows through the multicell RNN, the internal state of the cells changes and goes as feedback input for the next iteration (this is the key of memory). After that, there is a layer with a classical linear regression with the trainable variables W and b. 

Using the output of the linear regression and the encoded Y, the network calculates the cross-entropy and send it to a reduce sum layer, where it calculates the final loss. Using an Adam optimiser, this last layer back-propagates the error through all the layers (blue lines in fig. \ref{fig:rnnarchitecturetraining}) until reaching back the embeddings block. This procedure allows us to improve the embedding encoder and decoder, the cells coefficients, and the linear regression simultaneously.

In the case of sampling mode, the starting point is the seed, which is the sequence that the network will use to initiate its internal state and then generate new notes. From here, sampling mode has two phases. In the first one (green lines in fig. \ref{fig:rnnarchitecturesampling}), each sample of the seed flows through the embedding encoder to become its vector representation and then change the internal state of the cells in multicell RNN. Once it finishes with all the samples in seed in iteration m, phase 2 begins (blue lines in fig. \ref{fig:rnnarchitecturesampling}). The output of the multicell flows through the linear regression and a softmax layer which outputs the most probable next embedding. This output goes back as an input to the multicell RNN in the next iteration, and simultaneously it is decoded and added to the generated melody. This process repeats n times.

\subsection{Mono-midi-transposition-dataset} \label{subsec:dataset}

The dataset\footnote{The general dataset is available in \url{https://sebasgverde.github.io/mono-midi-transposition-dataset/}} used in this paper is a processed version of the mono-musicxml-dataset \cite{DBLP:journals/corr/WelU17}. The process consists of 3 steps\footnote{All the code, files and final datasets to replicate the results of the whole paper are available in \url{https://sebasgverde.github.io/rnn-cells-music-paper/}}: i) Scraping: The structure of the original dataset is used as the base to download the files from the web. ii) Prepossessing: Each midi is transformed into an array containing the sequence of notes (fig \ref{fig:dataset}.2). iii) Cleaning: only songs with more than 3 notes are kept (fig \ref{fig:dataset}.3).  The final list of arrays is used as the base to create 3 datasets (fig \ref{fig:dataset}.4 and \ref{fig:dataset}.5): 

\begin{figure}[h!]
    \centering
    \includegraphics[width=\textwidth]{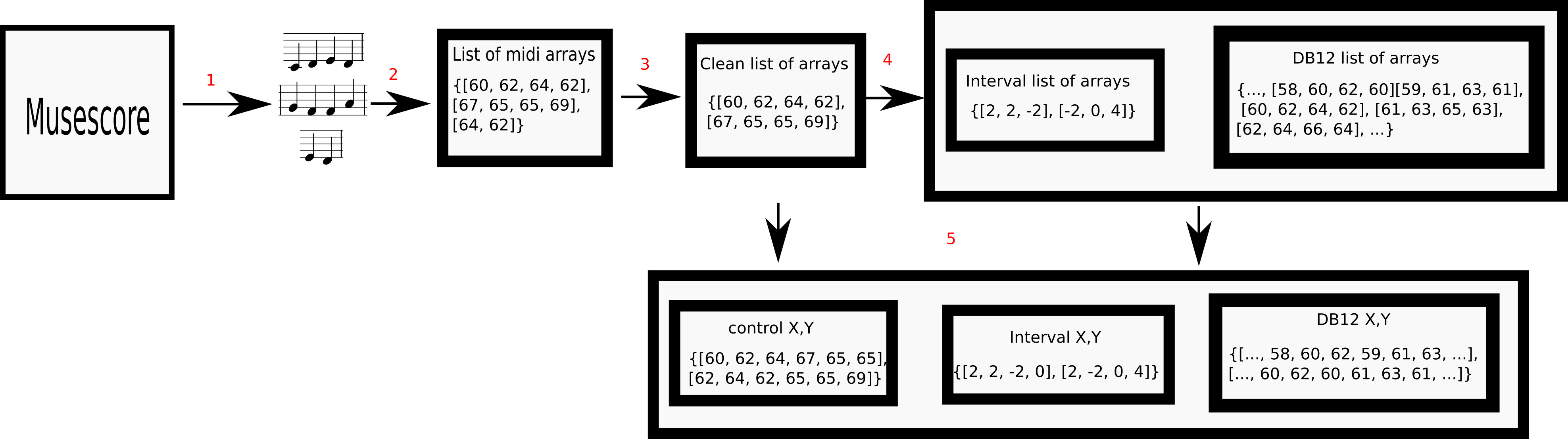}
    \caption{Dataset workflow}
    \label{fig:dataset}
\end{figure}

\textbf{Control Dataset: }This is the base case, on which X is the concatenated array of the original songs and Y is X shifted by one position.

\textbf{DB12 Dataset: }To construct this dataset, each song is transposed 12 times (on each degree of the chromatic scale). Pseudocode \ref{alg:db12} summarises the procedure, defining a song as an ordered set of midi notes of the form \{$X_{0}, X_{1}, ..., X_{|song|}$\}, where $X_{i}$ is the note in the position i, with $X_i \in [0,127]$.  The general idea is to spread the song through the midi range (0-127). The first part of the algorithm calculates the middle point of the song and subsequently moves the song up and down by semitones depending on how far is from central c4 (60). The algorithm intentionally gives preference to high notes which are more common in melodies. The DB12 dataset is the equivalent of the data augmentation regularly done in computer vision to increase the number of images by rotations, translations and mirroring \cite{Overett2008, Ciresan2013}


\algnewcommand{\IfThenElse}[3]{
  \State \algorithmicif\ #1\ \algorithmicthen\ #2\ \algorithmicelse\ #3}
  
\begin{algorithm}
\caption{Converting one song in 12}\label{alg:db12}
    \begin{algorithmic}[1]
    \Procedure{songToDB12}{$song$}
     \State $centralC\gets 60$
     \State $minNote\gets min(song)$
     \State $maxNote\gets max(song)$
     \State $middlePoint \gets \lfloor \frac{maxNote - minNote}{2}\rfloor+minNote$
     \State $centralCgap\gets centralC-middlePoint$
     \State $remainingTransp\gets 11 - \left | centralCgap \right |$
     
    \If  {$remainingTransp \geq 0$}
        \State $up\gets \left \lceil \frac{remainingTransp}{2} \right \rceil$
        \State $down\gets remainingTransp - up$
        \If {$centralCgap < 0$}
            \State $down\gets down + \left | centralCgap \right |$
        \Else
            \State $up\gets up +  centralCgap $
        \EndIf
    \Else
        \IfThenElse{$centralCgap \leq 0$}
        {$\text{down} = 11$}
        {$\text{up} = 11$}
    \EndIf
    
    \State $db12List\gets song$
    
    \For{$i \in [0, down]$}
    \State $newSong\gets \emptyset$
    \For{$X \in song$}
    \State $newSong[i]\gets X-(i+1)$
    \EndFor
    \State $db12List\gets db12List \cup newSong$
    \EndFor
    
    \For{$i \in [0, up]$}
    \State $newSong\gets \emptyset$
    \For{$X \in song$}
    \State $newSong[i]\gets X+(i+1)$
    \EndFor
    \State $db12List\gets db12List \cup newSong$
    \EndFor
    
     \State \textbf{return} $db12List$
    \EndProcedure
    \end{algorithmic}
\end{algorithm}

\textbf{Intervals Dataset: }In this case, we do not have a sequence of notes, but a sequence of relative changes. The idea of having a representation of the transitions between notes in a melodic context is not new (\cite{Hornel1998, Kaliakatsos-Papakostas2010, Johnson2017}). In a classification problem, the authors in \cite{Chai2001} shows that using relative changes improves the efficiency of their algorithm with respect to other approaches. Pseudocode \ref{alg:interval} shows our strategy to convert each song in an interval representation.

    
    

\begin{algorithm}
\caption{Converting one song in an interval array}\label{alg:interval}
    \begin{algorithmic}[1]
    \Procedure{songToInterval}{$song$}
    
    \State $intervalArray\gets \emptyset$
    
    \For{$X_i \in song $ with $i \in [0,|song|-1]$}
    \State      $intervalArray[i]\gets X_{i+1} - X_i$
    \EndFor
    \State \textbf{return} $intervalArray$
    \EndProcedure
    \end{algorithmic}
\end{algorithm}

\subsection{Description of the Quantitative Evaluation} \label{subsec:quantmetricdescr}

The quality of a musical piece is a subjective matter, and it is usually one of the most challenging parts of automatic music composition,  A common strategy to quantify the quality of a composition is to use human evaluations. However, the seminal work \cite{Tymoczko2011} proposes five music characteristics based on geometry (i.e. Conjunct Melody Motion, Acoustic Consonance, Harmonic Consistency, Limited Macroharmony and Centricity).
Together, these characteristics contribute to the tonality and the perceived beauty of a piece of music. Only 3 of this five features apply to melodies (i.e. Conjunct Melody Motion, Limited Macroharmony and Centricity). Acoustic consonance and Harmonic consistency apply exclusively to harmonies, and according to Tymoczko, these are the most culturally specific. These two features are not considered in this study. In the next subsections, the basic description of these features and their implementation are briefly introduced\footnote{The metrics are proposed only for songs with at least 12 notes.}.

Following the previously introduced notation of a song as an array, Let's define the $span_{i}$ as a subset of song from $X_{i}$ to $X_{i+n}$, where n = $|$span$|$ is the size of the span (In this case the constant 12), and  $Sq$ the quantity of spans given by equation \ref{eq:equationquantityspans}.

\begin{equation}
    Sq= 
\begin{cases}
    1, & \text{if } |song| \leq n\\
    |song|-n+1, & \text{otherwise}
\end{cases}
\label{eq:equationquantityspans}
\end{equation}

\subsubsection{Conjunct Melodic Motion (CMM)}
CMM looks for smooth transitions through the melody, in other words, the changes between notes must have short intervals. To capture this, we sum all the changes in the notes and divide it by the number of notes of the song minus 1. In our case the CMM is given by equation \ref{eq:cmm}. In general, the CMM is 1 if every change is of 1 semitone, tends to be greater than one, if there are changes of more than one semitone, and less than 1 for melodies with repeated notes. The best CMM is one because it is a song in the chromatic scale and sounds consonant by defect.

\begin{equation}
    CMM = \frac{1}{|song|-1} \sum_{i=1}^{|song|-1} \left |X_{i+1} - X_i \right |
    \label{eq:cmm}
\end{equation}

\subsubsection{Limited Macroharmony (LM)}

Macroharmony measures the diversity of notes in melodies. This is captured by measuring the number of different notes in short slices of time (spans).  For explanatory purposes, lets divide the Limited Macroharmony into two concepts: 

\begin{itemize}
\item{\textbf{Local Limited Macroharmony ($llm$)} quantifies the diversity of notes in a particular span. According to Tymoczko, tonal melodies must have between 5 and 8 different notes since the number of alternatives a human can manage is $7\pm2$ \cite{Miller1956}. Using a span size of 12 causes that the maximal penalisation for notes out the range (5,8) is the same if the span has either 1 or 12 different notes. It is important to remember that time is out of the scope of this paper and all the songs are always quarter notes so in this case, even when the book defines the span regarding musical time, in our case it reduces to the number of notes. 

The $llm$ is given by equation \ref{eq:llm}, where $lb$ is the lower bound, $ub$ is the upper bound and $d_{j}$ is the number of distinct notes in the particular span. In our case $lb$ = 5 and $ub$ = 8.  The $llm$ is 1 if the number of different notes in the span matches the appropriate range. Otherwise, it is the difference between the number of different notes in the span and the appropriate range plus 1.  \footnote{Since the same note in different octaves like D4 and D5 are different alternatives, we treat them just as different notes.}

\begin{equation}
    llm= 
\begin{cases}
    1,& \text{if } lb\leq d_{j} \leq ub\\
    (lb-d_{j})+1, & \text{if } dj < lb \\
    (d_{j}-ub)+1,  & \text{otherwise}
\end{cases}
\label{eq:llm}
\end{equation}
}

\item{\textbf{Limited Macroharmony ($LM$)} measures the diversity in the complete melody.  For melodies longer than the span size, the $lm$ is the average of calculating the $llm$ over a sliding window. The $lm$ is given by Equation \ref{eq:lm}, where the span quantity $Sq$ is the number of sliding windows covering a melody. In this case $lm$ is 1 if every span in a melody is within the appropriate limits, or greater than one for songs with a large diversity of notes.

\begin{equation}
    LM = \frac{1}{Sq} \sum_{j=1}^{Sq} llm_{j}
    \label{eq:lm}
\end{equation}
}
\end{itemize}

\subsubsection{Centricity (CENTR)}

Centricity claims that in short slices of time (spans), there must be a note which appears with more frequency than the others. We calculate the frequency of every note as the number of occurrences. Centricity is the max ratio between each note frequency in a span and the total notes in that span. To calculate the centricity of the complete song, we use the same strategy of average over a sliding window form LM. CENTR is given by equation \ref{eq:centr}, where i is defined in the interval $[0, ... ,  d_{j}]$ and $f(X_{ij})$ is the  frequency of note i in span j. In general CENTR tends to 1 the more predominant is one of the notes with respect to the others and tends to 0 the less frequent is the most frequent note.

\begin{equation}
    CENTR = \frac{1}{Sq} \sum_{j=1}^{Sq} max_i(\frac{f(X_{ij})}{n})
    \label{eq:centr}
\end{equation}

\subsubsection{Geometry descriptive statistics for dataset}

To properly evaluate the models, it is essential to define a baseline. The natural step is to apply the proposed metrics to the original dataset. Since the other 2 datasets are transformations of the original, the tonality in the songs should not differ. As table \ref{tab:metricDataset} shows, for the CMM and the LM the average is above the perfect score of 1 by 1.42 and 0.67 respectively, this evidence that a score above 1 in the generated pieces is not necessary a bad result. In the case of centricity, we can conclude that a tonal song should have a note that is at least the 20\% of the melody.

\begin{table}[h!]
\centering
\begin{tabular}{ccc} \hline
CMM     & LM      & CENTR \\ \hline
2.42 $\pm$ 0.98 & 1.67 $\pm$ 0.86 & 0.20 $\pm$ 0.12 \\
\hline
\end{tabular}

\caption{Dataset evaluation}\label{tab:metricDataset}
\end{table}


\section{Experiments} \label{sec:experiments}

This section compares different types of RNN cells (3), with different depth (5) and different transposition strategies (3). The combination of these decisions gives 45 different models to be analysed. This section defines the optimal number of layers, for every data-set and memory cell, reducing the number of models to 9 (3 datasets and 3 memory cells). Looking for simplicity the experiments are grouped by data-set and the convergence of the cross-entropy used as evaluation metric. 

Regarding the training conditions: i) In all the cases the batch size and the sequence length is 50, which means that every training iteration uses 50 sequences of 50 notes simultaneously. ii) For the control and the interval dataset, the model trains 300 epochs, what means that uses the complete dataset 300 times. For the DB12 dataset, which is 12 times bigger than the other 2, it trains only 50 epochs. 

\subsection{Control dataset experiment}

The optimal number of layers for the three cell types is 3. The worst performance is for 1 and then 5 layers, this is interesting since it shows that this is not a simple matter of add as much layers as possible.

\begin{figure}[h!]
    \centering
    \begin{subfigure}[b]{0.4\textwidth}
        \includegraphics[width=\textwidth]{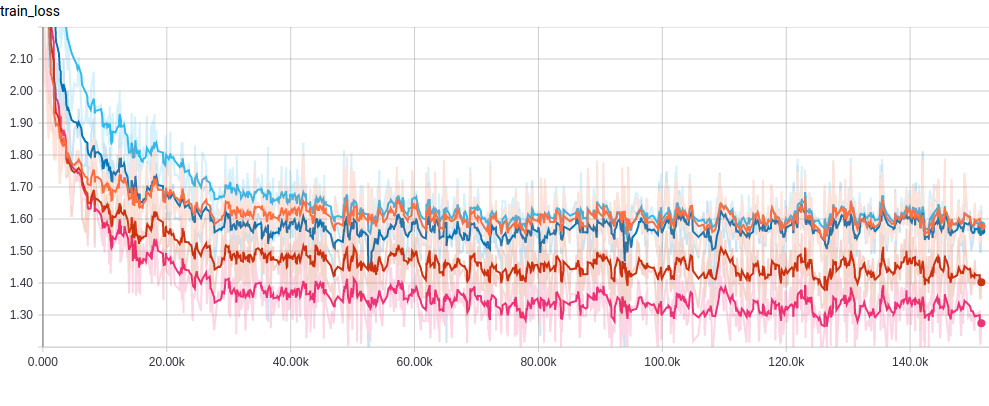}
        \caption{LSTM}
        \label{fig:controllstm}
    \end{subfigure}
    
    \begin{subfigure}[b]{0.4\textwidth}
        \includegraphics[width=\textwidth]{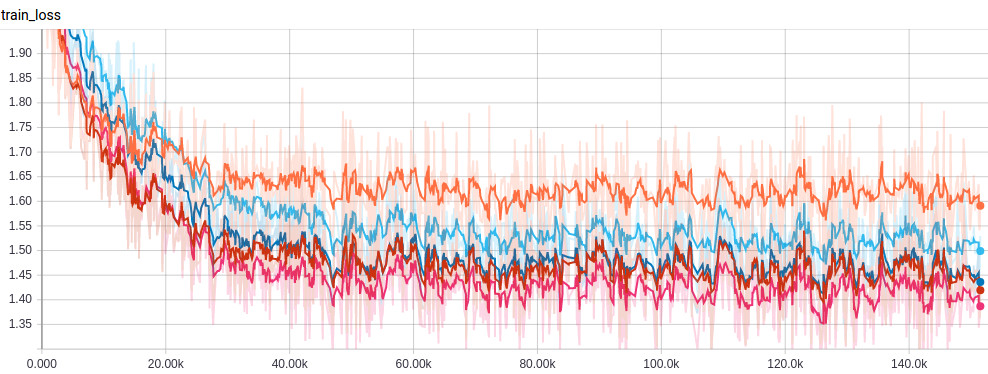}
        \caption{NAS}
        \label{fig:controlnas}
    \end{subfigure} \\
    
    \begin{subfigure}[b]{0.4\textwidth}
        \includegraphics[width=\textwidth]{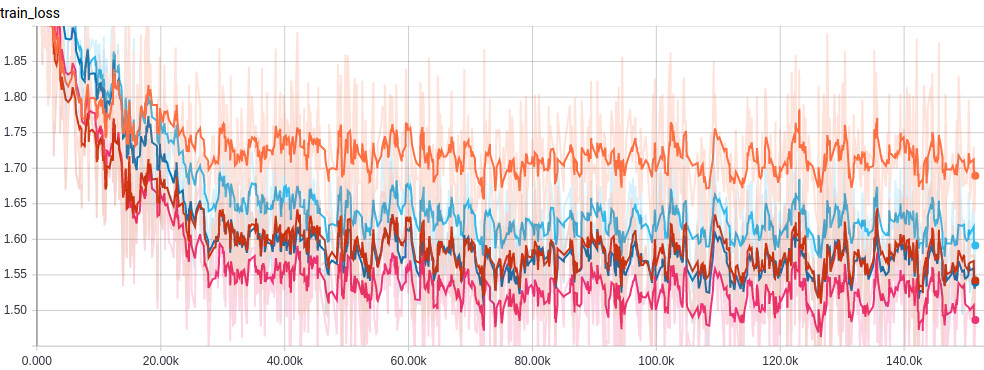}
        \caption{UGRNN}
        \label{fig:controlugrnn}
    \end{subfigure}
    
    [\textcolor{orange}{\textbullet}] 1 layer [\textcolor{red}{\textbullet}] 2 layers [\textcolor{magenta}{\textbullet}] 3 layers [\textcolor{blue}{\textbullet}] 4 layers [\textcolor{cyan}{\textbullet}] 5 layers
    \caption{Control Dataset Experiment Learning Curves}\label{fig:control}
\end{figure}

The five-layer model tends to outperform the one-layer one after some iterations. This happens around 50k for LSTM  (Fig. \ref{fig:controllstm}) and 20k for NAS  (Fig. \ref{fig:controlnas}) and UGRNN  (Fig. \ref{fig:controlugrnn}). The four and five-layer models always begin with a worse score, what means that the more layers, the more it takes to converge.

\subsection{Interval dataset Experiment}

\begin{figure}[h!]
    \centering
    \begin{subfigure}[b]{0.4\textwidth}
        \includegraphics[width=\textwidth]{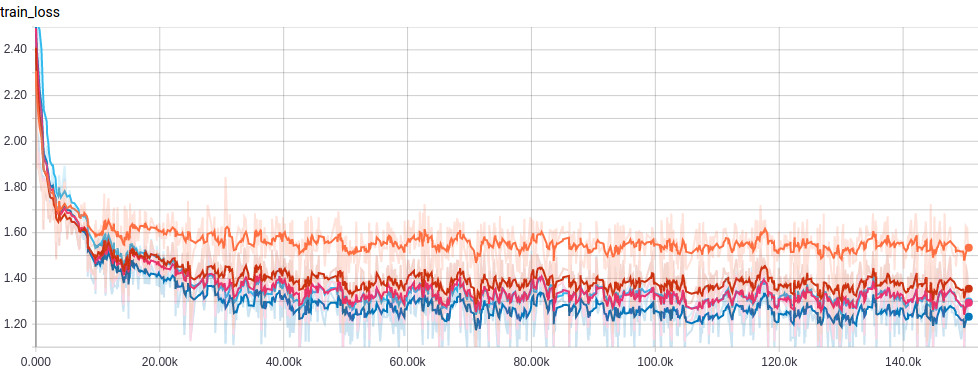}
        \caption{LSTM}
        \label{fig:intervallstm}
    \end{subfigure}
    
    \begin{subfigure}[b]{0.4\textwidth}
        \includegraphics[width=\textwidth]{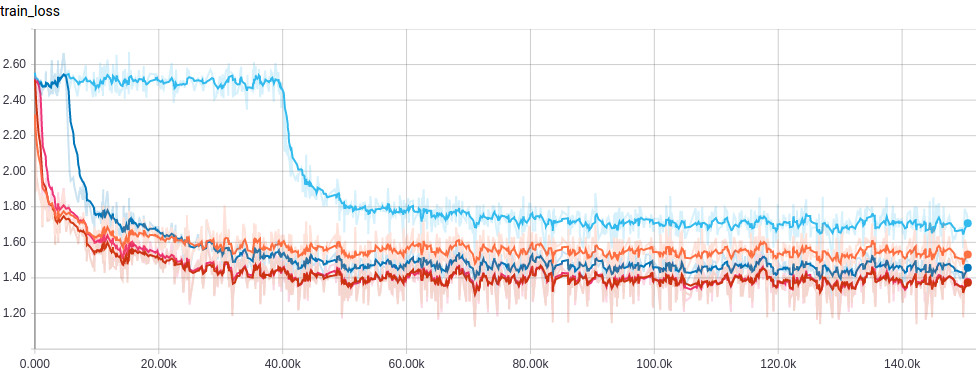}
        \caption{NAS}
        \label{fig:intervalnas}
    \end{subfigure} \\
    
    \begin{subfigure}[b]{0.4\textwidth}
        \includegraphics[width=\textwidth]{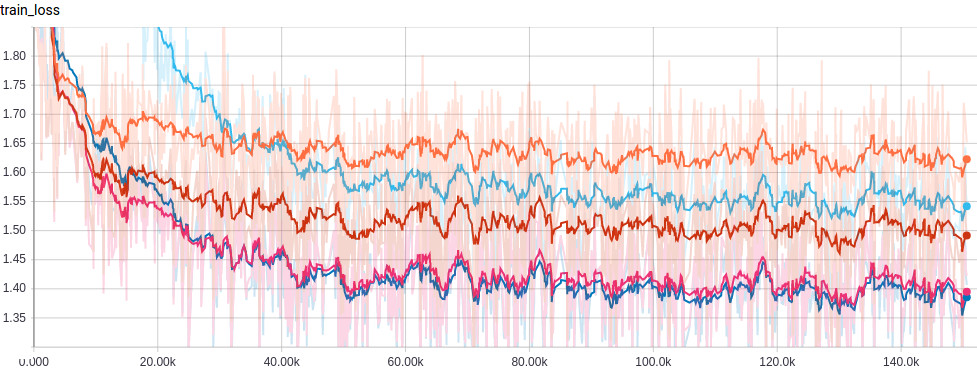}
        \caption{UGRNN}
        \label{fig:intervalugrnn}
    \end{subfigure}
    
    [\textcolor{orange}{\textbullet}] 1 layer [\textcolor{red}{\textbullet}] 2 layers [\textcolor{magenta}{\textbullet}] 3 layers [\textcolor{blue}{\textbullet}] 4 layers [\textcolor{cyan}{\textbullet}] 5 layers

    \caption{Interval dataset Experiment Learning Curves}\label{fig:interval}
\end{figure}

Both, the four and five-layer models outperform the one and two-layer after approximately 17k iterations with LSTM cells (Fig. \ref{fig:intervallstm}), in this case, 4 is the optimal number of layers at the end.

For the NAS (Fig. \ref{fig:intervalnas}), the most in-depth models take a considerable number of iterations to begin to converge, this is mainly evident for 5 layers, which only improve after 40k iterations. These deep models, however, never outperform the two-layer one, which is the best in this case. 

The UGRNN experiment (Fig. \ref{fig:intervalugrnn}) shows an interesting behaviour. The five-layer model outperforms the one-layer after 35k iterations, these two models are at the end the fourth and fifth place respectively. The four-layer model outperforms all the others gradually, the one-layer at 15k, two-layer at 19k and three-layer at 40k, until it reaches the lowest loss at the end

\subsection{DB12 Dataset Experiment}

\begin{figure}[h!]
    \centering
    \begin{subfigure}[b]{0.4\textwidth}
        \includegraphics[width=\textwidth]{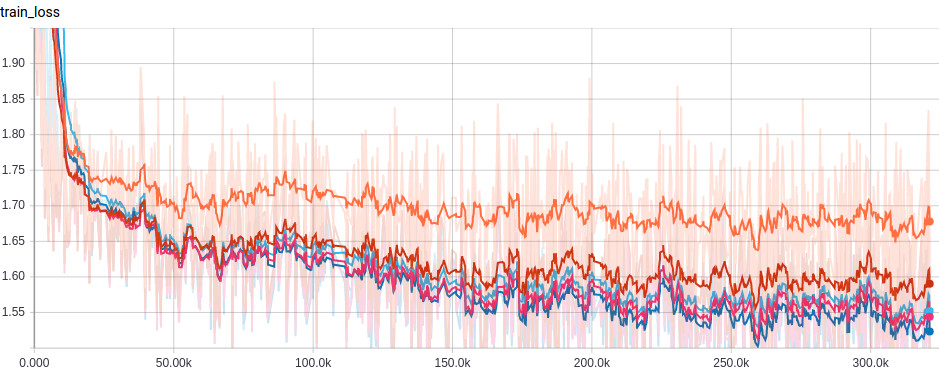}
        \caption{LSTM}
        \label{fig:db12lstm}
    \end{subfigure}
    
    \begin{subfigure}[b]{0.4\textwidth}
        \includegraphics[width=\textwidth]{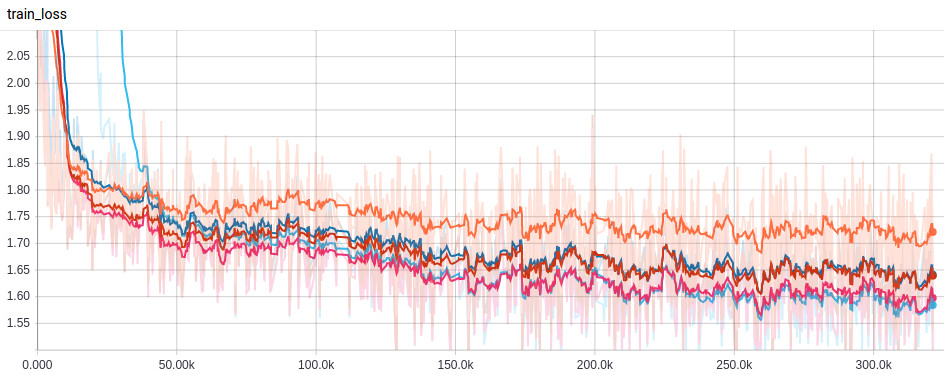}
        \caption{NAS}
        \label{fig:db12nas}
    \end{subfigure} \\
    
    \begin{subfigure}[b]{0.4\textwidth}
        \includegraphics[width=\textwidth]{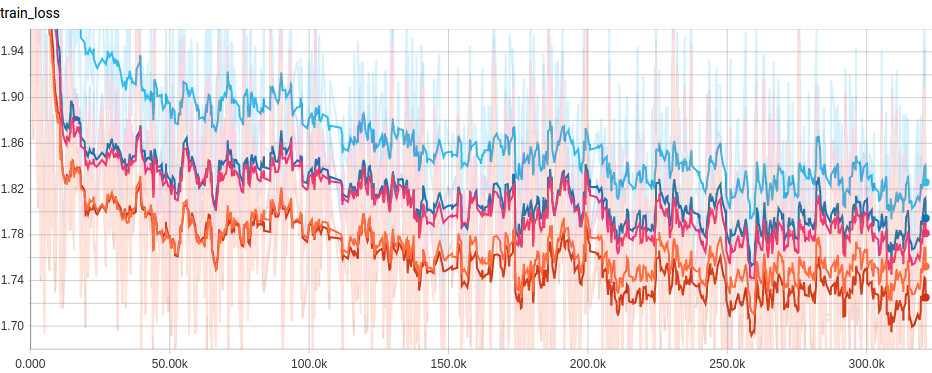}
        \caption{UGRNN}
        \label{fig:db12ugrnn}
    \end{subfigure}
    
    [\textcolor{orange}{\textbullet}] 1 layer [\textcolor{red}{\textbullet}] 2 layers [\textcolor{magenta}{\textbullet}] 3 layers [\textcolor{blue}{\textbullet}] 4 layers [\textcolor{cyan}{\textbullet}] 5 layers

    \caption{DB12 Dataset Experiment Learning Curves}\label{fig:db12}
\end{figure}

The LSTM (Fig. \ref{fig:db12lstm}) and NAS (Fig. \ref{fig:db12nas}) have a similar behaviour since the worst result in both cases is for 1 and 2 layers. The LSTM experiment (Fig. \ref{fig:db12lstm}) shows that 4 is the number of layers that reach the best result. In the NAS experiment on the other hand (Fig. \ref{fig:db12nas}), the five-layer model begins as the worst, after iteration 50k, it gradually outperforms the others and finally reaches the first place.

In UGRNN experiment (Fig. \ref{fig:db12ugrnn}), results are stable from the beginning until the end. The five-layer model is the worst, while one and two layer models are the best. This cell type is the only one with a good performance for 1 layer, however, is the two-layer model which has the best loss.

\subsection{Analysing the Network Depths}

In general, the learning curves presented in previous sections show that the deeper the network, the more iterations it needs to converge. Table \ref{tab:optimalnumberoflayers} summarises the optimal number of layers for each dataset and memory mechanism showing no consensus for all the cases and making necessary a more detailed analysis.  The following part of this section analyses the advantages and disadvantages of each dataset (rows) and the performance of each memory-mechanism (columns). The remaining part of this paper uses the number of layers defined in Table \ref{tab:optimalnumberoflayers}.

\begin{table}[h!]
\centering
\begin{tabular}{lccc} \hline
         & LSTM     & NAS      & UGRNN \\ \hline
CONTROL  & 3 & 3 & 3 \\
INTERVAL & 4 & 2 & 4 \\
DB12     & 4 & 5 & 2 \\
\hline
\end{tabular}

\caption{Optimal number of layer per model}\label{tab:optimalnumberoflayers}
\end{table}


Regarding the datasets, it is interesting, that the \textit{Control dataset} is the only one with no variation in the optimal number of layers for the three cells, suggesting it as the most stable dataset. It is the opposite case for the \textit{DB12}, which is the dataset where the LSTM and UGRNN (the similar memory mechanism) diverge. Additionally, \textit{DB12} shows the higher variance concerning the depth; it is the only dataset where the deepest models are the best option (DB12-NAS models) and the only one with a good performance for a one layer architecture according to the convergence curve (see fig. \ref{fig:db12ugrnn}). These facts suggest the \textit{DB12 dataset} as the less stable of the three datasets. 

Regarding the type of memory mechanism, the LSTM and UGRNN (which have an established and quite similar architecture) show a similar behavior, except for the less stable DB12 dataset. The nature of the NAS cell, which does not have a defined architecture but builds it while training, combined with the representation of intervals seems to cause a divergence with LSTM and UGRNN. Finally, this evolutionary nature of the NAS cell, combined with the less stable DB12 dataset, can be the cause of the five layers result.

\section{Results} \label{sec:results}

This chapter introduces the evaluation and analysis of the 9 models of table \ref{tab:optimalnumberoflayers}. In section \ref{subsec:automEval100songs} the 3 metrics described in \ref{subsec:quantmetricdescr} test 100 songs for each of the 9 models to quantify their composition capabilities. Section \ref{subsec:generatedmelodies}, visualises the 9 most representative songs, and provide analysis from the musical viewpoint.

\subsection{Analysis of Metrics in the Models}\label{subsec:automEval100songs}

To evaluate the 9 models from table \ref{tab:optimalnumberoflayers}, each model generates 100 songs with the same parameters: the sequence C4-D4-E4-D4 as seed and 30 new notes. Table \ref{tab:metricsSelectedModel} shows the average and standard deviation for each metric and model. Concerning the best model for each metric, it is not possible to conclude due to the magnitudes of the standard deviations; However, under the light of each metric, important insights can be obtained.

\begin{table}[h!]
\centering
\begin{tabular}{clccc} \hline
      &  & LSTM     & NAS      & UGRNN \\ \hline
\multirow{ 3}{*}{CMM} &  
CONTROL  & 2.14 $\pm$ 0.60 & 1.90 $\pm$ 0.49 & 2.11 $\pm$ 0.50 \\
&INTERVAL  & 2.99 $\pm$ 1.08 & 2.44 $\pm$ 1.00 & 2.42 $\pm$ 0.74 \\
&DB12  & 2.49 $\pm$ 0.59 & 4.28 $\pm$ 2.71 & \textbf{1.66 $\pm$ 0.40} \\
\hline
\multirow{ 3}{*}{LM} &  
CONTROL  & \textbf{1.47 $\pm$ 0.46} & 1.77 $\pm$ 0.65 & 1.51 $\pm$ 0.46 \\
&INTERVAL  & 2.37 $\pm$ 1.56 & 1.59 $\pm$ 0.75 & 1.66 $\pm$ 0.84 \\
&DB12  & 1.57 $\pm$ 0.64 & 2.56 $\pm$ 1.11 & 2.40 $\pm$ 0.65 \\
\hline
\multirow{ 3}{*}{CENTR} &  
CONTROL  & 0.20 $\pm$ 0.08 & 0.22 $\pm$ 0.09 & 0.21 $\pm$ 0.09 \\
&INTERVAL  & 0.16 $\pm$ 0.08 & 0.16 $\pm$ 0.09 & 0.17 $\pm$ 0.12 \\
&DB12  & 0.16 $\pm$ 0.08 & 0.25 $\pm$ 0.26 & \textbf{0.23 $\pm$ 0.09} \\
\hline

\end{tabular}

\caption{Means and standard deviation for the 100 songs generated by the final models with the 3 metrics.}\label{tab:metricsSelectedModel}
\end{table}


According to the Conjunct Melodic Motion (CMM) the closest CMM to 1 is the \textit{DB12-UGRNN} (1.66), with a considerable advantage over \textit{Control-NAS}(1.90), and outperforming the average of the dataset (2.42). It is interesting how the \textit{DB12-NAS} model shows a significantly worse result, and the largest standard deviation, positioning it as the least reliable model according to the CMM.

In the case of LM the control dataset have the best performance for this metric with 3 of the best results. Regarding, to the cell type, each has a good performance with 2 datasets and a bad one with the other, giving a result of 3 wrong combinations: INTERVAL-LSTM, DB12-NAS and DB12-UGRNN. The model with best LM is the CONTROL-LSTM with 1.47, what also outperforms the 1.67 of the dataset (table \ref{tab:metricDataset}), while the worse is the DB12-NAS. 

Regarding the centricity, The model which shows higher centricity is the \textit{DB12-NAS}, however, the bad performance of this model in the other metrics indicates a strong repetition of notes. Also, it has the highest standard deviation, therefore, the \textit{DB12-UGRNN} with only 0.09 of deviation and just 0.2 less centricity is the best candidate. Notice that this both models have poor LM, this makes sense, while centricity measures the prevalence of a note, the LM, makes the opposite.

In general, LSTM models are stable through the datasets for CMM and shows a low centricity (tends to use more notes), what is evident in the good results for LM. The NAS cell seems to be unstable with the DB12 dataset. Between the 2 strategies for transposition learning, the interval representation is the most stable. 

Respect to the best model for the 3 metrics, the \textit{CONTROL-NAS}, model has a good result for all the metrics. It is the only model that outperforms the centricity of the dataset without affecting the LM considerably, which is the weakness of the \textit{DB12-UGRNN} model. In summary, concerning only the quantitative analysis, the best models are i)  \textit{DB12-UGRNN} for CMM and CENTR, ii) \textit{CONTROL-LSTM} for LM and iii) \textit{CONTROL-NAS} for general tonality.

\subsection{Generated Melodies Musical Analysis} \label{subsec:generatedmelodies}

Using the average metrics for the 100 songs generated by each of the 9 models (table \ref{tab:metricsSelectedModel}), it is possible to select a representative song per model using the Euclidean Distance to the average as selection criteria. Table \ref{tab:metricsMostRepresSongs} summarises the metrics for sampled songs.

\newcommand{\rotationtalbeangle}{60}
\setlength{\tabcolsep}{3pt}
\begin{table}[h]
\centering
  \begin{tabular}{lccccccccc}
\cline{2-10}
 & \multicolumn{3}{c}{LSTM} & \multicolumn{3}{c}{NAS} & \multicolumn{3}{c}{UGRNN} \\\cline{2-10}           
 & \rotatebox[origin=c]{\rotationtalbeangle}{CMM}& \rotatebox[origin=c]{\rotationtalbeangle}{LM}& \rotatebox[origin=c]{\rotationtalbeangle}{CENTR}    
 & \rotatebox[origin=c]{\rotationtalbeangle}{CMM}& \rotatebox[origin=c]{\rotationtalbeangle}{LM}& \rotatebox[origin=c]{\rotationtalbeangle}{CENTR} 
 & \rotatebox[origin=c]{\rotationtalbeangle}{CMM}& \rotatebox[origin=c]{\rotationtalbeangle}{LM}& \rotatebox[origin=c]{\rotationtalbeangle}{CENTR} \\
 \hline
CONTROL  & 2.21 & 1.52 & 0.17 & 1.85 & 1.78 & 0.20 & 2.06 & 1.39 & 0.26 \\
INTERVAL  & 2.73 & 2.22 & 0.10 & 2.36 & 1.65 & 0.24 & 2.42 & 1.70 & 0.18 \\
DB12  & 2.45 & 1.61 & 0.08 & 4.48 & 2.70 & 0.29 & 1.58 & 2.39 & 0.25 \\
\hline
\hline
\end{tabular}

\caption{The 3 metrics for the most representative of the 100 songs generated by the final models in each case.}\label{tab:metricsMostRepresSongs}
\end{table}

\begin{figure}[h]
    \centering
    \begin{subfigure}[b]{0.45\textwidth}
        \includegraphics[width=\textwidth]{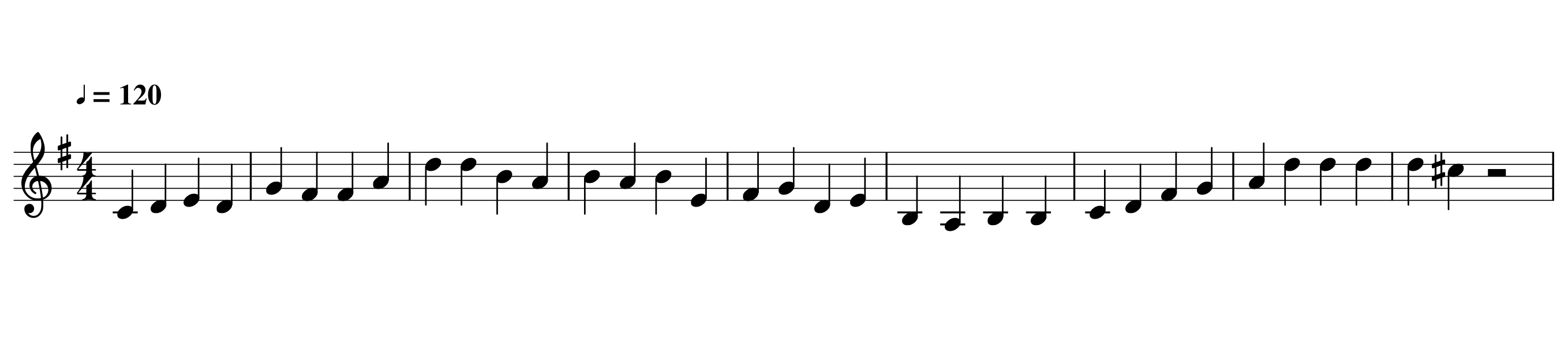}
        \caption{LSTM Model}
        \label{fig:baselstmsong}
    \end{subfigure}
    ~ 
    \begin{subfigure}[b]{0.45\textwidth}
        \includegraphics[width=\textwidth]{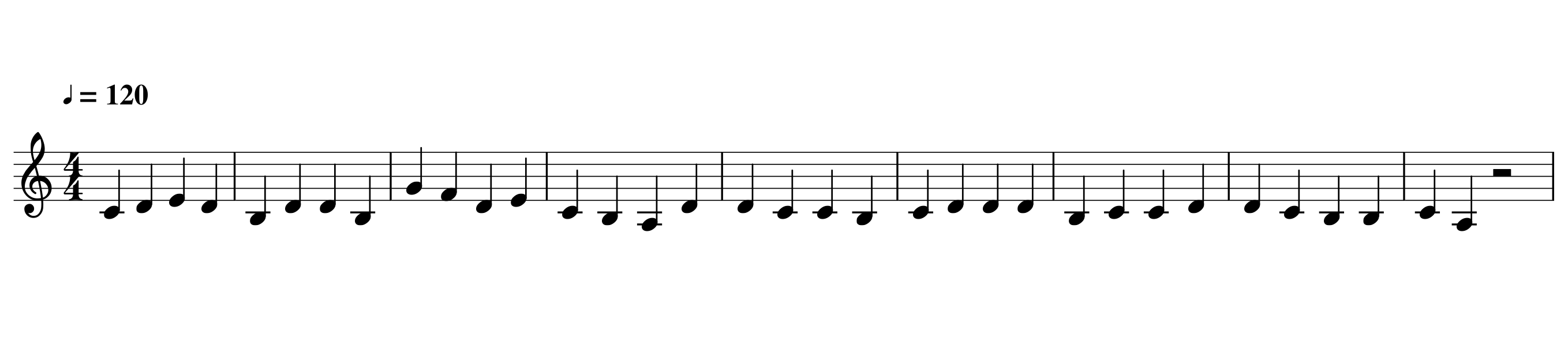}
        \caption{NAS Model}
        \label{fig:basenassong}
    \end{subfigure} \\
    ~ 
    \begin{subfigure}[b]{0.45\textwidth}
        \includegraphics[width=\textwidth]{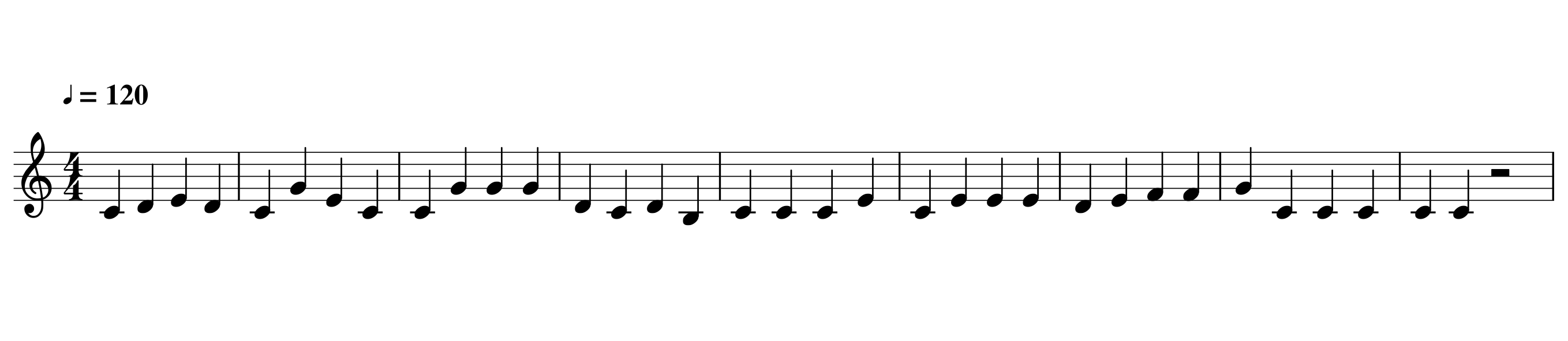}
        \caption{UGRNN Model}
        \label{fig:baseugrnnsong}
    \end{subfigure}
    \caption{Generated Melodies Control Dataset}\label{fig:generatedmelodiescontrol}
\end{figure}

\begin{figure}[h]
    \centering    
    \begin{subfigure}[b]{0.45\textwidth}
        \includegraphics[width=\textwidth]{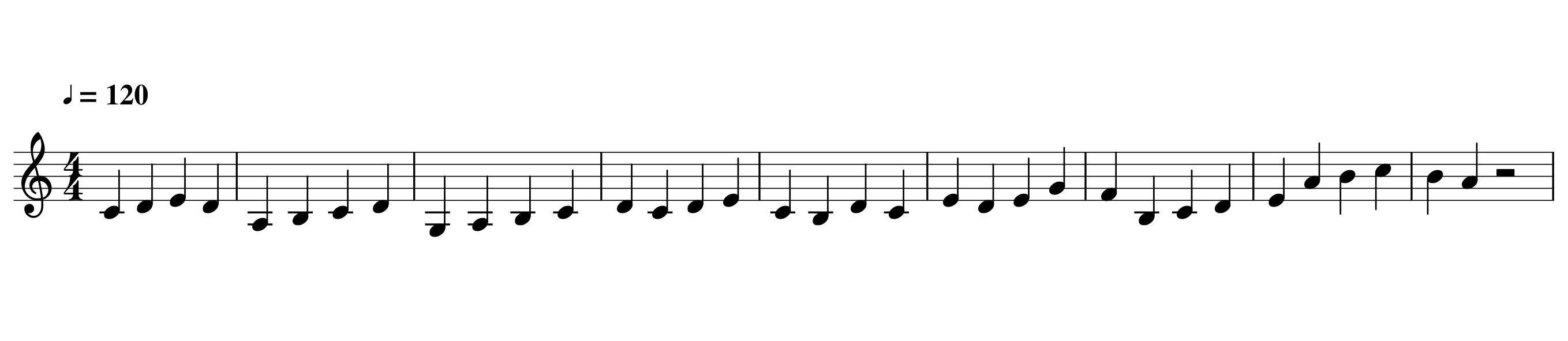}
        \caption{LSTM Model}
        \label{fig:db12lstmsong}
    \end{subfigure}
    ~ 
    \begin{subfigure}[b]{0.45\textwidth}
        \includegraphics[width=\textwidth]{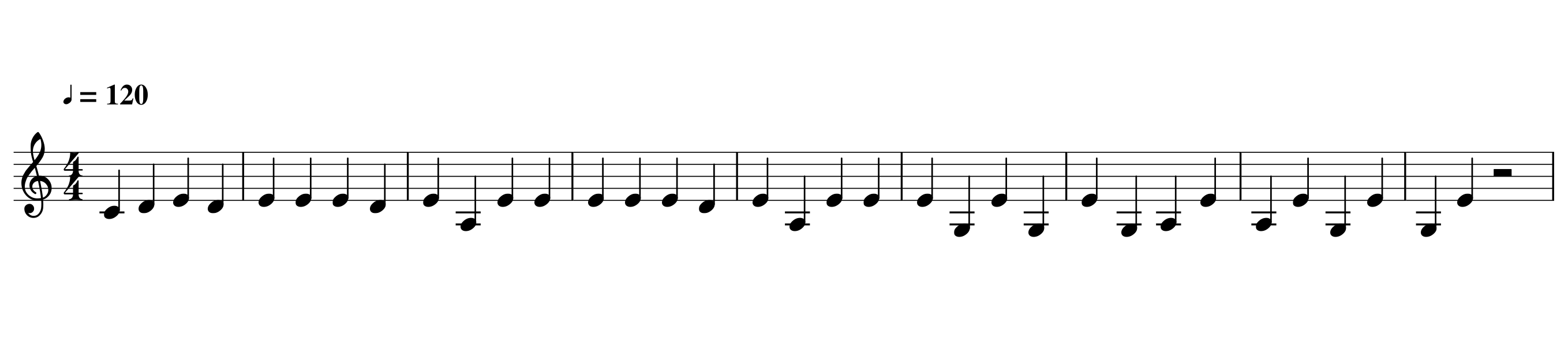}
        \caption{NAS Model}
        \label{fig:db12nassong}
    \end{subfigure} \\
    ~ 
    \begin{subfigure}[b]{0.45\textwidth}
        \includegraphics[width=\textwidth]{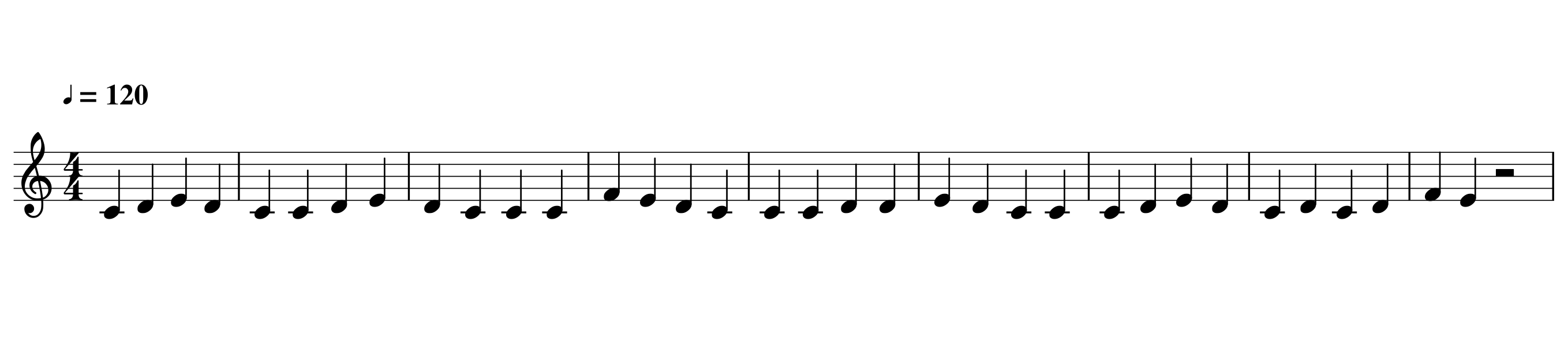}
        \caption{UGRNN Model}
        \label{fig:db12ugrnnsong}
    \end{subfigure}    
    \caption{Generated Melodies DB12 Dataset}\label{fig:generatedmelodiesdb12}
\end{figure}

\begin{figure}[h!]
    \centering    
    \begin{subfigure}[b]{0.45\textwidth}
        \includegraphics[width=\textwidth]{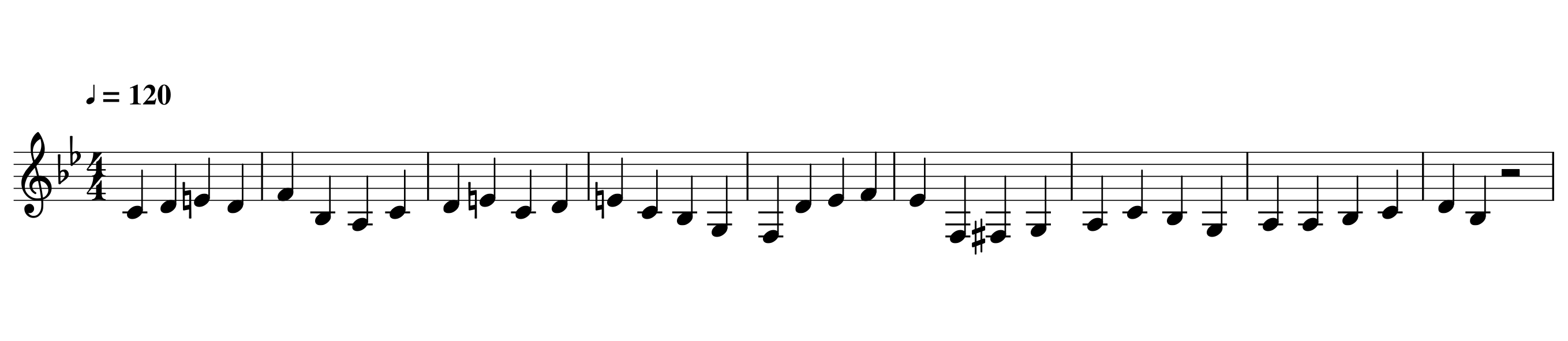}
        \caption{LSTM Model}
        \label{fig:intervallstmsong}
    \end{subfigure}
    ~ 
    \begin{subfigure}[b]{0.45\textwidth}
        \includegraphics[width=\textwidth]{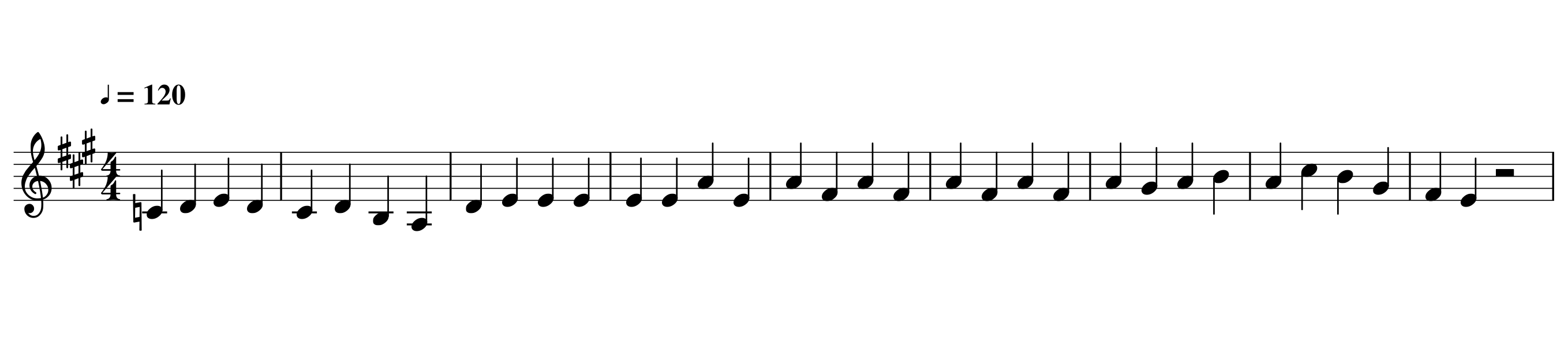}
        \caption{NAS Model}
        \label{fig:intervalnassong}
    \end{subfigure} \\
    ~ 
    \begin{subfigure}[b]{0.45\textwidth}
        \includegraphics[width=\textwidth]{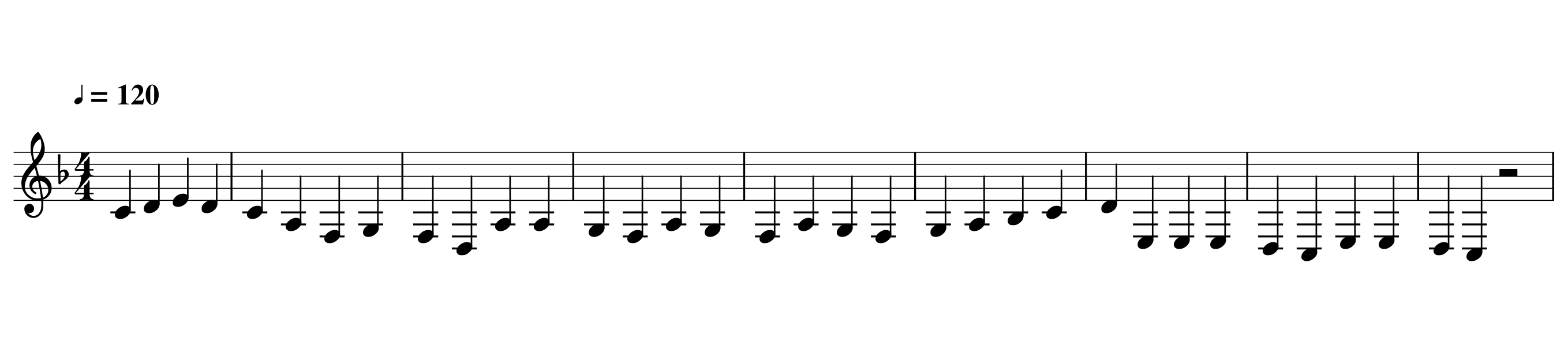}
        \caption{UGRNN Model}
        \label{fig:intervalugrnnsong}
    \end{subfigure}    
    \caption{Generated Melodies Interval Dataset}\label{fig:generatedmelodiesinterval}
\end{figure}

Figures \ref{fig:generatedmelodiescontrol}, \ref{fig:generatedmelodiesdb12} and \ref{fig:generatedmelodiesinterval} shows the musical sheet of the 9 songs\footnote{Listen to the 9 songs in \url{https://youtu.be/FGUIEshh6WU}} from table \ref{tab:metricsMostRepresSongs}. The CONTROL-LSTM model (fig. \ref{fig:baselstmsong}) has a balanced number of notes. It also shows an ascendant melody which goes from C4 to A4 in bar 7. In the case of the NAS (fig. \ref{fig:basenassong}) and UGRNN (fig. \ref{fig:baseugrnnsong}) variations, the first shows a descendant pattern in bar 3 going from G4 to D4. The second, as expected from the metrics for UGRNN, shows a repetition of notes, but in this case, it forms a tonal pattern, repeating groups of 3: G4 in bar 3, C4 in bar 5 and E4 in bar 6; these are the notes that harmonise the C major chord. Under the light of current results it is not clear if the 5 notes repeated at the end, are the result of a stagnant of the model.

As expected by the metrics (Tables \ref{tab:metricsSelectedModel} and \ref{tab:metricsMostRepresSongs}), the DB12-NAS model (fig. \ref{fig:db12nassong}) tends to repeat the notes, in this case, using the E4 19 times, and in the middle of the song 5 times consecutively, this naturally favors the centricity. The other model with good centricity DB12-UGRNN (fig. \ref{fig:db12ugrnnsong}) also repeats, in this case, the note C4, but in this case more tonally. The good performance of CMM is evident, unlike DB12-NAS which at the end has changes of even 9 semitones (major sixth), this model keeps softer changes.

Finally, it is important to speak about the models which did not highlight in the metrics evaluation. The DB12-LSTM (fig. \ref{fig:db12lstmsong}) shows the most interesting ascendant patterns in the 9 songs, going from A3 to D4, then changing to G3, to go to C4, and at the end goes from B3 to resolve in C5. Notice that there are not accidentals in this melody, the model does not know by representation, the difference between natural, sharp and flats, they are just midi numbers. It could learn this patterns of the C major scale.
In the case of the interval dataset, the LSTM (Fig. \ref{fig:intervallstmsong}) shows many changes and never repeat consecutively notes, this was expected for the low centricity and high CMM. the UGRNN (Fig. \ref{fig:intervalugrnnsong}) seems not to explore many changes in the pattern, and the NAS (Fig. \ref{fig:intervalnassong}) even if at the beginning of bar 5 seems to be stagnant in a pattern A4-F♯4, in bar 7 it resolves, and makes a melody with clear use of A4 as pedal note.

\section{Conclusions and future research} \label{sec:conlusions}

About the 3 cell types, it is interesting that in the representative melodies of the models, the consecutive repetition of notes in LSTM is almost null, especially with the DB12 dataset, this cell shows an excellent learning capacity to keep the scale. The NAS model shows in general a high standard deviation for the metrics when trains with DB12, NAS models need many iterations to begin to converge (Sec \ref{sec:experiments}), the combination of this aspect with the more significant size of DB12 dataset results in a less stable model. Finally, melodies from UGRNN models shows a good capacity to use consecutive notes as root pattern for the melodies.

In the musical context, the depth of the net is not a trivial problem of using so many layers as possible, section \ref{sec:experiments} shows many examples where after 2 or 3 layers, the cost function is worse. Therefore, it is necessary to make always tests with gradual increments in the layer number to check the optimal.

The automatic metrics are a useful tool to see the general behavior of a model, they can describe it and potentially be the reward to train the models using a strategy like reinforcement learning. However, the 3 metrics used in the paper only describe tonality,  they does not have into account other aspects like rhythm, timbre, performance or rubato. They are not enough to make a full judgment of a melody, the best proof of this are some of the songs (fig. \ref{fig:intervalnassong}, \ref{fig:db12lstmsong}), which even when their models do not have a highlight performance in the comparison table \ref{tab:metricsSelectedModel}, sound tonal and have interesting patterns.

With respect to the dataset variations, the control dataset is the most stable for the LM metric, the interval dataset shows a low centricity, and even when none of the interval models had a highlight result in the metrics, its scores are stable among them. The db12 variation models on the other hand, show a high centricity and tends to repeat the notes a lot, but also, the embeddings shows a better understanding of the notes musical meaning, and in general seems to capture better the patterns than the interval variation. A DB12 representation tends to be more complicated than an interval like one, as result, interval has a more stable behaviour in the metrics (Table \ref{tab:metricsSelectedModel}), however, if the purpose is to have a transposition free representation of musical patterns, section \ref{subsec:generatedmelodies} shows DB12 in combination with embedding encoding as the best alternative.

Finally, the general results of the trained models shows that cross-entropy is as valid as the objective function in the musical context as it has been through the years for machine learning models in all the areas where it has been applied.

As future work, the natural step will be to extend the model to support rhythm and harmonies. It is relevant to see the potential of embeddings in music, our results in this research show learning of particular features of the notes as pitch level and note nature (fundamental or altered). In this vein, a broader study, including harmonies and time changes can show more appealing patterns, potentially, a trained embedding can cluster the musical elements in genres, having applications in tasks like genre classification. It would also be interesting to see the performance of our optimal nine models when used to generate sequences in other areas as i) Text: poetry or story writing, using the words which a poem should have, or the main idea of the story as initialisation. ii) Audio: audio synthesis and voice generation.
Finally, the implementation of the automatic metrics is a good starting point to make them more robust and include others applicable to rhythm and harmony. The inclusion of qualitative evaluation to compare and refine the metrics can also improve its capacity to measure music quality. These metrics could be used not only to measure the quality and tonality of the final generated songs but to train the model using reinforcement learning to optimise the metrics directly.

\section{Acknowledgement}

Special thanks to the Center of Excellence and Appropriation on Big Data and Data Analytics (Alianza CAOBA) and to the APOLO high performance computing centre for the GPU hardware provided, which was very helpful to reduce the time of training.

\section{Funding}

This research was supported by the Center of Excellence and Appropriation on Big Data and Data Analytics (Alianza CAOBA).

\section{Supplementary Materials}
As support for the paper, there is a web page in the url: \url{https://sebasgverde.github.io/rnn-cells-music-paper/} with all the material necessesary for research replication, which includes:

\begin{enumerate}
    \item Datasets
    \item Network Weights
    \item Midi songs of the final models
    \item Link to demo videos
    \item RNN model
    \item Library for midi manipulation
    \item Library for music evaluation
    \item Detailed instructions to run the scripts
\end{enumerate}

Scripts available include:
\begin{enumerate}    
    \item Environment setting
    \item Datasets and weights downloading
    \item Datasets unit tests
    \item Layer experiment
    \item Generating songs with trained models
    \item Music Evaluation of models, songs and datasets
    \item Embeddings experiment 
    \item Cross-entropy Validation
\end{enumerate}

\bibliographystyle{apalike}
\bibliography{bibliography.bib}
\end{document}